\documentclass[runningheads]{llncs}
\usepackage[T1]{fontenc}
\usepackage{graphicx}
\usepackage{hyperref}
\usepackage{color}

\usepackage{annotate-equations}
\usepackage{amsmath}
\usepackage{amssymb}

\usepackage{url}
\usepackage{booktabs}
\usepackage{multirow}
\usepackage{makecell}
\usepackage[inline]{enumitem}
\usepackage[subtle, mathspacing=normal]{savetrees}
\newcommand{\vspaceSecBefore}[0]{\vspace{-.7em}}
\newcommand{\vspaceSecAfter}[0]{\vspace{-.3em}}
\newcommand{\vspaceSSecBefore}[0]{\vspace{-.5em}}
\newcommand{\vspaceSSecAfter}[0]{\vspace{-.2em}}
\begin{document}
\title{The Cost of Executing Business Processes\\ on Next-Generation Blockchains:\\ The Case of Algorand}
\titlerunning{Executing Processes on Next-Gen Blockchains}
%\author{Ran Jecker\inst{1}\orcidID{0000-1111-2222-3333} \and
%J. aus Brocken\inst{2,3}\orcidID{1111-2222-3333-4444} \and
%Mr. Becker\inst{3}\orcidID{2222--3333-4444-5555}}
%
\authorrunning{Stiehle and Weber}
\author{Fabian Stiehle\inst{1} \and Ingo Weber\inst{1,2}}
%\author{Authors Redacted}
%}
%
%\authorrunning{Ivar Ekeland et al.}   % abbreviated author list (for running head)
%Craig Chambers, Kim B. Bruce, Elisa Bertino}
%
\institute{%\email{fabian.stiehle@tum.de, ingo.weber@tum.de}
Technical University of Munich, School of CIT, Germany, \email{first.last@tum.de} \and
    Fraunhofer Gesellschaft, Munich, Germany
}
%\institute{for Double Blind Review}
%
\maketitle              % typeset the header of the contribution
\begin{abstract}
Process (or workflow) execution on blockchain suffers from limited scalability; specifically, costs in the form of transactions fees are a major limitation for employing traditional public blockchain platforms in practice. Research, so far, has mainly focused on exploring first (Bitcoin) and second-generation (e.g., Ethereum) blockchains for business process enactment. However, since then, novel blockchain systems have been introduced---aimed at tackling many of the problems of previous-generation blockchains. We study such a system, Algorand, from a process execution perspective. Algorand promises low transaction fees and fast finality. However, Algorand's cost structure differs greatly from previous generation blockchains, rendering earlier cost models for blockchain-based process execution non-applicable. We discuss and contrast Algorand's novel cost structure with Ethereum's well-known cost model. 
% drawing from insights from Ethereum's evolution from a relatively static cost model to a more dynamic at present. 
To study the impact for process execution, we present a compiler for BPMN Choreographies, with an intermediary layer, which can support multi-platform output, and provide a translation to TEAL contracts, the smart contract language of Algorand. We compare the cost of executing processes on Algorand to previous work as well as traditional cloud computing. In short: they allow vast cost benefits. However, we note a multitude of future research challenges that remain in investigating and comparing such results. 
\keywords{Blockchain,
Process Execution,
Process Enactment,
Workflow,
Choreography}
\end{abstract}
\vspaceSecBefore
\section{Introduction}
\vspaceSecAfter
Blockchain can serve as a decentralised platform to ensure the enforcement of process rules and the visibility and integrity of execution traces. 
The current state of the art executes processes or workflows either through a smart contract that interprets some representation of the process or by transforming a process model into a smart contract implementation~\cite{stiehle2022blockchain}. The focus of most earlier works has been on investigating what can be achieved based on so-called first (i.e., Bitcoin) and second-generation blockchain systems (providing turing complete smart contracts)~\cite{xu2017taxonomy,stiehle2022blockchain}. Due to their public and decentralised nature, they have to employ mechanisms to incentivise honest verifiers (e.g., miners) and protect against denial-of-service and Sybil attacks~\cite{tschorschBitcoinTechnicalSurvey2016}. This necessitates paying fees for transactions. Early work investigating these platforms for process execution demonstrate that these fees are considerable, especially compared to traditional cloud computing~\cite{prybilaRuntimeVerificationBusiness2020a,rimbaQuantifyingCostDistrust2020a}. With the increasing popularity of these platforms, this problem has since become more pronounced.
Stiehle and Weber~\cite{stiehle2022blockchain} % the authors  
trace the historic development of transaction fees on public Ethereum and conclude that these have become prohibitively expensive for most process cases~\cite[p. 16]{stiehle2022blockchain}. To achieve better performance, for a small set of process participants, private blockchain deployments can be employed. However, % Stiehle and Weber note, 
many of the guarantees a blockchain can offer are a result of decentralisation (see~\cite{stiehle2022blockchain} or~\cite[Chapter~3.2]{xuArchitectureBlockchainApplications2019}). Certain high-security use cases may benefit from larger, more decentralised, and resilient deployments~\cite{stiehle2022blockchain,kjaer2023towards}.
Thus, in~\cite{stiehle2022blockchain} the authors chart two directions for improving the status-quo: exploring (i) novel next-generation blockchain platforms, and (ii) layer two solutions.
Early results on the use of layer-two solutions show that these can considerably reduce transaction fees~\cite{stiehle2023processchannels}. However, these solutions build on top of existing blockchain platforms (layer one) and add additional layers of complexity and trade-offs.

So far, blockchain-based cost models are primarily investigated from a protocol perspective, e.g., to identify certain attack vectors~\cite{perez2019broken}.
Only few works exist from a system design perspective~\cite{rimbaQuantifyingCostDistrust2020a,pincheira2023infrastructure}. These models have a heavy focus on Ethereum's cost structure~\cite{rimbaQuantifyingCostDistrust2020a} or take an infrastructure perspective, i.e., they remain abstract on the underlying model~\cite{pincheira2023infrastructure}.
We fill a gap in literature, by providing a formalisation and discussion of the cost model of Algorand, and contrast it to the well-studied cost model of Ethereum, which is the basis of existing related work.
Our study of Algorand is motivated by its promise of fast finality and low cost~\cite{gilad2017algorand}, as well as---compared to Ethereum---opposing design decisions and qualities. Examples for the latter are Algorand's fixed cost and minimum balance requirement system.
To operationalise our formalisations for the use of process execution, we extend an open source compiler for BPMN Choreographies towards multi-platform support and Algorand specifically. To the latter end, we provide a translation from choreographies to TEAL contracts, the native smart contract language of Algorand. 

Here, we discuss different design options for TEAL contracts, with different cost implications as indicated by our cost model analysis. 
We also present an efficient multi-instance execution support.
We evaluate our cost model, and compare our design to previous work as well as traditional cloud computing, utilizing previous results of Rimba et al.~\cite{rimbaQuantifyingCostDistrust2020a}.
Our results highlight the promise of executing processes on next-gen blockchains---in short: they allow vast cost benefits. However, we note the many challenges that remain in investigating and comparing such results. 
Specifically, our investigation stands as early result showing that investigating novel blockchain protocols is valuable, yet by no means trivial. Many new characteristics must be considered by application developers. For organisations, predictable cost and performance estimates are paramount---we see our study as first contribution to a more holistic understanding. Thus, we believe that our results extend beyond the narrow domain of process execution on blockchain.
Following open science principles, we release our implemented artefact as open source software.\footnote{\label{fn:artefact}\url{https://github.com/fstiehle/chorpiler-algorandvm}. An archived version is available at: \url{https://doi.org/10.5281/zenodo.12656684}.}

\vspaceSecBefore
\section{Background}
\vspaceSecAfter
We assume some familiarity with the Ethereum protocol and focus on introducing Algorand. Nonetheless, we will contrast concepts of Algorand to Ethereum, and retrace certain, more recent developments of Ethereum's evolution; specifically, on its cost model. 
%For space reasons, we focus on major updates since the work of Rimba et al.~\cite{rimbaQuantifyingCostDistrust2020a} in 2018.
In the remainder of this manuscript, we discuss cost as consequence of transacting on the public blockchain. Primarily, this is expressed by fees or balance requirements.

%\subsection{A Short History of Ethereum's Cost Model}
%The evolution of a blockchain protocol usually requires hard forks, as verifiers need to upgrade to the new protocol rules. In Ethereum, such forks implement a range of Improvement Proposals (EIP), in Figure, we give an overview of forks implementing EIPs that affect the cost model. We focus here on the most radical ones: 
%We give a more detailed analysis of the current state in Section~\ref{sec:cost-model}.

\vspaceSSecBefore
\subsection{Algorand}
\vspaceSSecAfter
\subsubsection{Consensus.}
The Algorand protocol relies on verifiable random functions (VDF), which produce a randomness and proof based on a secret key and seed~\cite{gilad2017algorand}. The proof enables other participants to verify that the randomness is indeed the result of running VDF with seed and secret key. 
%The seed is determined for each new block. 
Each proposer in the network locally runs VDF with a seed determined for each new block and their key. Given that their result is within a given threshold they are a candidate to produce the next block. To achieve scalable byzantine fault tolerance, the VDF is also run to determine membership in a committee. This committee constitutes a random sample of the network, and consensus is executed within this sample. To prevent targeted attacks on committee members (once they are elected), committee members reveal their membership only by participating in the protocol once, after, a new committee is formed. In the common case, the proposed block is finalised and no valid fork can exist in the network. Some edge cases exist where Algorand favours safety over liveness, and additional steps are required. For more details we refer the reader to~\cite{gilad2017algorand}. To protect against Sybil attacks, the VDF is weighed by stake (i.e., proof-of-stake).
\vspace{-.9em}
\subsubsection{Accounts, Contracts, and Transactions.}
Algorand\footnote{We describe Algorand based on the Algorand developer documentation (\url{https://developer.algorand.org/docs/get-details}, accessed 2024-07-01), but also based on our practical experiments; specifically when documentation remained too ambiguous for our purposes.} accounts can be broadly divided into four types: (i) external accounts, (ii) multisignature accounts, (iii) application accounts, and (iv) smart signature accounts. 
All accounts can be associated with a balance of \textit{algo} (ALG), assets, and local application state.

Smart contracts fall into two categories: stateful applications and stateless signatures. Signatures can not hold state (but the corresponding account can hold assets, ALG, and local state) and are attached to a transaction. Applications can hold state through different storage types. Throughout Algorand's development, applications became more and more potent, making the use of signatures only necessary for some niche scenarios.\footnote{\url{https://developer.algorand.org/docs/get-details/dapps/smart-contracts/guidelines/\#more-about-smart-signatures-vs-smart-contracts}, accessed 2024-07-01.} We also focus on applications for the remainder of this paper.
%Algorand transactions can be related to payments (with ALG), asset management, registration for participation in the consensus, and applications. 
An application is created through a transaction and requires the specification of an \verb|approval| and \verb|clear| program in the programming language TEAL, as well as the allocation of the required static storage type. Hereby, the approval program implements the main application logic. An application call succeeds when the \verb|approval| program returns with a non-zero value on top of its stack. Otherwise, the transaction is rejected. The \verb|clear| program handles deletion of local state and cannot fail. Beyond the standard application call type, the app can be called by update, delete, and local state transactions. Through local state transactions, accounts can allocate local state within an application (associated with their account). This state can be cleared at any time by the associated account. An application can, in-turn submit transactions of any type, called inner-transactions.
Atomic transaction groups can batch together transactions to be executed in an all-or-nothing fashion. 
%This can also be used to increase these limitations of a singular transaction.
Applications can use different storage schemes. 
Through an associated application account, they can hold ALG and assets. Global and local storage are fixed and must be allocated during application creation. 
%Local storage, however, is dynamic in the sense of the amount of accounts that opt-in. That is, local storage is assigned per account (would be more clear when explained in terms of key, account, value… maybe a table would be worth compiling?). 
Only box storage can be allocated on-demand by an application. We analyse these storage systems in detail in Section~\ref{sec:cost-model}.
\vspaceSecBefore
\section{Related Work}
\vspaceSecAfter
Cost models in blockchain are often studied from a protocol perspective, where incentives (e.g.,~\cite{abbasi2022algorand}) or attack vectors, based on a mismatch of computational effort and cost, are investigated, e.g.,~\cite{perez2019broken}. Another line of research is focused on explaining and predicting certain market determinants, e.g.,~\cite{donmez2022transaction}. Cost models from a system design perspective are less investigated, specifically for next-generation blockchains with novel cost structures. Pincheira et al.~\cite{pincheira2023infrastructure} provide a  model from a high abstraction level, to gauge the infrastructure cost of operating an application on a public vs. private blockchain, which they evaluate based on an Ethereum environment.

From a process perspective, Kj{\"a}er et al.~\cite{kjaer2023towards} provide a qualitative discussion of different blockchain technology groups. Our work can be seen as a concrete, primarily quantitative, investigation of the transaction costs concern for a new technology group. Rimba et al.~\cite{rimbaQuantifyingCostDistrust2020a} provide a concrete model of the Ethereum cost structure. Our choice of Algorand for this investigation is motivated by the many opposing qualities of its cost model compared to the much studied Ethereum. Furthermore, since Rimba et al., Ethereum's cost model has evolved in many ways, we will review.

From the perspective of multi-platform support for process execution, there is work towards enabling the generation and composition of multiple blockchain target implementations~\cite{ladleifArchitectureMultichainBusiness2020,corradiniModeldrivenEngineeringMultiparty2021a,falaziProcessbasedCompositionPermissioned2019}. 
However, their focus is on second-generation and private blockchain targets. Furthermore, the underlying cost model is not explored. Many aspects of these works are complimentary to our work, such as the  blockchain independent access layer described in Falazi et al.~\cite{falaziProcessbasedCompositionPermissioned2019}.

In~\cite{xu2022distributed}, Xu et al. propose a method to translate DCR graphs to TEAL contracts. In contrast, we study a petri-net based translation technique as proposed in~\cite{garcia-banuelosOptimizedExecutionBusiness2017a}, but adapted for BPMN Choreographies. More importantly, our study also differs in scope. We give a formalised treatment of Algorand's underlying cost model, based on which we are able to discuss different design options. We present an architecture for a multi-target compiler, which supports cost efficient execution of multiple instances, and contrast our results to Ethereum's cost model and traditional cloud computing.
%However, they do not discuss the underlying cost model in detail, nor different design options are a consequence of it. 
\vspaceSecBefore
\section{Preliminary Cost Model Analysis}
\label{sec:cost-model}
\vspaceSecAfter
Algorand's cost model incorporates two aspects: (i) transaction fees, which incur directly for every transaction; and, (ii) minimum balance requirements, which is a complete separate notion, and incur whenever storage slots on the chain are allocated. Balance requirements are not cost \textit{per se}, as it encodes an amount of ALG an account must hold at a minimum.

To study Algorand's cost model in more detail, we introduce Ethereum as a baseline, as the most influential and, arguably, \textit{archetype} of second-gen blockchains. Contrasting new developments against this archetype (and all its problems) can help classifying and contextualising them. Specifically, as Ethereum has come under the most scrutiny---especially for process execution~\cite{stiehle2022blockchain}. 
%Current cost models are heavily based on this archetype~\cite{rimbaQuantifyingCostDistrust2020a}, but are incompatible with Algorand's design.
\vspaceSSecBefore
\subsection{Transaction Fees}
\vspaceSSecAfter
Transaction cost in Ethereum are a function of the gas units used $n_{\text{gas}}$ and the gas price, expressed in gwei ($10^{-9}$ ETH).
Since EIP-1559~\cite{buterin2019eip1559}, the gas price can no longer be freely set by a client. The price is now determined by the base fee $b_{\text{gas}}$, which increases or decreases based on the previous block's utilisation. 
%Eg., if the previous block was above the block size target, the base fee increases. 
A client can add a priority fee $p_{\text{gas}}$, to incentivise faster inclusion of the transaction. The verifier only receives $n_{\text{gas}} \times p_{\text{gas}}$ as a reward, the rest is made inaccessible (burned).
%Through this level of indirection, where some parts of the fee can be freely chosen, the convergence to a gas price exhibits qualities of an open market (TODO: Bad wording, CITE).
Transactions start with a base gas amount of $n_{\text{gas}}=21,000$. The additional amount is determined through execution, where each opcode is metered. 
Opcodes are priced based on their computational complexity and many exhibit dynamic cost, i.e., their cost depend on the current execution context, not just the opcode and its input. Formally, the transaction cost $c^{\text{tx}}_{\text{eth}}$ is
\begin{equation}
\label{eq:eth_tx}
c^{\text{tx}}_{\text{eth}} = n_{\text{gas}} \cdot  (b_{\text{gas}} + p_\text{gas}).
\end{equation}
%\annotate[]{below, left}{congeste}{congestion control}
%\annotate[]{below, right}{incentive}{incentive}

\noindent In Algorand, transaction cost is at least the fixed value $b_\text{alg} = 1,000$, here expressed in malgo ($10^{-6}$ ALG). Should the verifier be congested, the transaction size $s_{\text{tx}}$ becomes the determining factor. 
Formally, transaction cost $c^{\text{tx}}_{\text{alg}}$ is
\begin{equation}
\label{eq:alg_tx}
c^{\text{tx}}_{\text{alg}} = \text{max}(b_\text{alg},\, f_{\text{alg}} \cdot s_{\text{tx}}),
\end{equation}
%\annotate[]{below, left}{congesta}{congestion control}
%\vspace{1em} 
where $f_{\text{alg}}$ is calculated based on the local level of congestion of a verifier.\footnote{\url{https://github.com/algorand/go-algorand/blob/v3.24.0-stable/data/pools/transactionPool.go}, accessed 2024-06-04.} Thus, contrary to Ethereum, the client can not choose any parameter, congestion is calculated locally rather than globally, and there is no reward payed to the verifier for processing the transaction. Furthermore, there is no incentive the client can provide for their transaction to be prioritised (cf.~\cite{oz2024study}).

Algorand's transaction fee do not directly account for computational complexity. Rather, a transaction is constrained by an opcode cost limit $l_{\text{op}} = 700$. Opcodes are metered; however most (83\%)\footnote{\url{https://developer.algorand.org/docs/get-details/dapps/avm/teal/opcodes/v10}, accessed 2024-07-05.} are priced with a flat fee of $1$, and cost is not influenced by the execution context, just the opcode and its input.
The opcode limitation can be lifted by batching transactions within an atomic transaction group, e.g., one can issue "dummy" transactions to increase the budget for another opcode heavy transaction in the group.
The cost of a transaction group is just the sum of the included transaction costs, i.e., for $n_{\text{tx}}$ transactions, the cost is $c^{\text{tx}}_{\text{alg}} \times n_{\text{tx}}$, where $n_{\text{tx}} \leq 16$.\footnote{See \url{https://developer.algorand.org/docs/get-details/parameter}, accessed 2024-06-04.}
Thus, for the perspective of application development, we can write the cost of a singular opcode $c^{\text{op}}_{\text{alg}}$ as
\begin{equation}
\label{eq:opc}
c^{\text{op}}_{\text{alg}} = c^{\text{tx}}_{\text{alg}} \cdot (1 + \lfloor{\frac{n_{\text{op}} - 1}{l_{\text{op}}}}\rfloor),
\end{equation}
%\vspace{5em}
where $n_{\text{op}}$ is the metered cost of opcodes used.
Applications can issue an additional $256$ inner transactions, increasing the pooled opcode limit further.% to $190,400$. 
%Outer tx can pay for inner.
% % Below is incorrect as outer tx can pay for inner tx, so no funding of account required (as we only talk about tx cost here, not mb, so creating boxes still requires funding an app account)
%Following seems to be wrong, tx can also pool their fees, so outer tx can pay for inner (Can we then also do this for box storage!? according to the comment not, but we need to test): However, the contract must pay the transaction fees of inner transactions, which requires a funded account. In Algorand, each account requires a minimum balance $mb$ of $100,000 * 10^{-6}$ algo. TODO: How to get this balance back? We will discuss the concent of minimum balance in more detail in the next Section.
%If the contract does not need to function as an account, $mb$ is not required. That is, Equation \ref{eq:opc} holds for $11,200 < op_l$. 
%\begin{equation}
%c^\text{op}_{\text{alg}} =
%\begin{cases}
%c^{\text{tx}}_{\text{alg}} + c^{\text{tx}}_{\text{alg}} * \lfloor\frac{n_{op}}{op_l}\rfloor & op_l \leq 11,200 \\
%100,000 + c^{\text{tx}}_{\text{alg}} + c^{\text{tx}}_{\text{alg}} * \lfloor\frac{n_{op}}{op_l}\rfloor & 11,200 < op_l \leq 190,400
%\end{cases}
%\end{equation}
%n_{gas}$ is influenced by intrinsic cost (e.g., fixed transaction cost, cost per byte, payed prior execution) and a dynamic cost influenced by memory size and storage access.\footnote{Yellow Paper, Section 9.2 and 6.2}
%Opcodes have an intrinsic fixed cost as in Algorand. However, if memory usage is increased due to an operation, this will also be priced.\footnote{Yellow Paper, Appendix G}
\vspaceSSecBefore
\subsection{Storage Fees}
\vspaceSSecAfter
For the purpose of process execution, we focus on persistent storage that can be written and read in the context of a contract call, and is under the control of the (orchestrating) application (layer one storage).
Pricing such persistent memory is especially delicate for blockchain platforms, as these grow the replicated storage and require expensive disk operations~\cite{perez2019broken}.
In Ethereum, where everything is priced based on metered opcodes, the virtual machine (EVM) defines the \verb|SLOAD| and \verb|SSTORE| opcodes, which load and store byte key value pairs. %of the the combined size of $2*32$B.
Following the introduction of EIP-292~\cite{buterin2020eip2929}, these opcodes are metered based on whether storage is accessed for the first time, so called \textit{cold} storage, or not, so called \textit{warm} storage. An access list keeps track of warm slots during the current execution context. A default transaction pre-loads the address of the transaction sender and recipient to the list. An optional transaction parameter can be supplied to specify elements to additionally pre-load. However, due to the current implementation of this list, there is no cost advantage for most contract transactions~\cite{heimbach2023dissecting}.
%\footnote{This, however, can be mostly seen as a consequence of the current implementation, and is projected to change in the future, as pre-loading storage enables a multitude of performance improvements, see EIP-3521.}
%\footnote{A cost advantage can be gained should at least $24$ storage slots be pre-loaded and accessed, as the implementation requires to load the contract address as well when loading storage slots. However, this address is already preloaded by default. See~\cite{heimbach2023dissecting} for a more detailed discussion.}
%Explain the fallacy (I can add the storage keys, but I additionally pay for the contract address, which is no benefit (as this access is not priced). Cite Wattenhofer and EIP trying to rectify this.
%Table~\ref{tab:ethereum} gives an overview of the gas price of contract creation, for warm, and cold storage. 
%On Ethereum, a funded account (holding ETH) is not priced. 
In EIP-3529~\cite{buterin2021eip3529}, refunds were  drastically reduced. Refunds were introduced to incentivise the economical usage of Ethereum's storage space; specifically for data no longer required. Freeing a contract slot is no longer compensated; freeing a storage slot reimburses $4,800$  gas, but only up to a maximum of $\frac{1}{5}n_{\text{gas}}$ of the current transaction.

Algorand has two different storage systems that can be used: global and box storage. Both store key value pairs, where the key is limited to $64$B. Global storage can be divided into \verb|Uint|, a $8$B unsinged integer value, and \verb|Byte| storage, a max. $128$B byte value. In total, global storage is limited to 64 slots (i.e., key value pairs) per application. Box storage allocates a variable size byte value up to a limit of $32$KB, the number of boxes is not limited.

Storage in Algorand is priced by the notion of minimum balance, which is defined per account and expresses the balance the account is required to hold at a minimum. This balance is practically frozen, but can be refunded (unfrozen) by freeing the associated storage slot again. The opcodes to store and access storage are also metered, but the cost is a negligible $n_{\text{op}} = 1$ for these operations. See Table~\ref{tab:algorand} for an overview. In contrast to Ethereum, a funded account requires a minimum balance also. 
Similarly to Ethereum's access list, Algorand has the notion of reference lists. A standard application call transaction pre-loads two references: the caller and the callee. Storage slots outside of the reference lists can not be accessed. The reference list is limited to eight additional entries per transaction. Analogously to the opcode limit, this limit can be increased by pooling transactions. 
If a contract address is included in the reference list, it also includes all global storage slots associated. Box slots must be included separately and for each $1$KB. E.g., with one transaction, one can access $64$ global storage slots, i.e., $8$KB, and $8$KB box storage.%\footnote{Say some words about other limitations, e.g., read access to global and boxes, what else?}.
\newcolumntype{L}[1]{>{\raggedright\let\newline\\\arraybackslash\hspace{0pt}}m{#1}}
\newcolumntype{C}[1]{>{\centering\let\newline\\\arraybackslash\hspace{0pt}}m{#1}}
\newcolumntype{R}[1]{>{\raggedleft\let\newline\\\arraybackslash\hspace{0pt}}m{#1}}
\renewcommand\theadfont{}
%\begin{table}[tb]
%\scriptsize
%\centering
%\begin{tabular}{rR{2.3cm}R{1.1cm}R{1.1cm}} 
%\toprule
% &  & \multicolumn{2}{c}{\,\,\,\,Storage (64B)} \\
%    \cmidrule{3-4}
% & Contract ($n_\text{B}$) & \makecell[r]{Warm} & \makecell[r]{Cold}  \\
%\midrule
%Cost (gas) \\
%    Alloc & $32,000 + \text{bc($n_\text{B}$)}$ & $20,000$ & $22,100$  \\
%    Update & - & $^{\S}2,900$ & $5,000$  \\
%    Load & - & $100$ & $2,100$  \\
%Refund (gas)
% & 0 & $^{\dagger}4,800$ & $^{\dagger}4,800$ \\
%\bottomrule
% \multicolumn{4}{l}{$^{\S}$$100$ for consecutive updates.} \\
% \multicolumn{4}{l}{$^{\dagger}$max $\frac{1}{5}\,n_{\text{gas}}$ of the current transaction.}\\
%\end{tabular}
%\caption{Overview of different storage cost in the EVM.}
%\label{tab:ethereum}
%\vspace{-2em}
%\end{table}
\begin{table}[tb]
\centering
\scriptsize
\begin{tabular}{rR{1.2cm}R{1.1cm}R{1.9cm}R{1cm}R{1cm}} 
\toprule
 & & & \multicolumn{3}{c}{Storage} \\
    \cmidrule{4-6}
 & \makecell[r]{\,\\Account} & \makecell[r]{Contract\\ (2048B)} & \makecell[c]{Box\\ ($n_\text{B}$)} & \makecell[c]{Byte\\ (128B)} & \makecell[c]{Uint\\ (72B)} \\
\midrule
Cost \\
    ($mb$) & 100,000 & 100,000 & $2500 + 400\,n_\text{B}$ & $50,000$ & $28,500$ \\
    ($op$) & - & - & 1 & 1 & 1 \\
Refund (of $mb$)
 & $100\%$ & $100\%$ & $100\%$ & $100\%$ & $100\%$ \\
Limit ($B$) & - & $8,192$ & - & $8,192$ & $4,608$ \\
\bottomrule
\end{tabular}
\caption{Overview of the cost of different storage types in Algorand.}
\label{tab:algorand}
\vspace{-2em}
\end{table}
\begin{figure}[t]
    \centering

    \includegraphics[width=.7\linewidth]{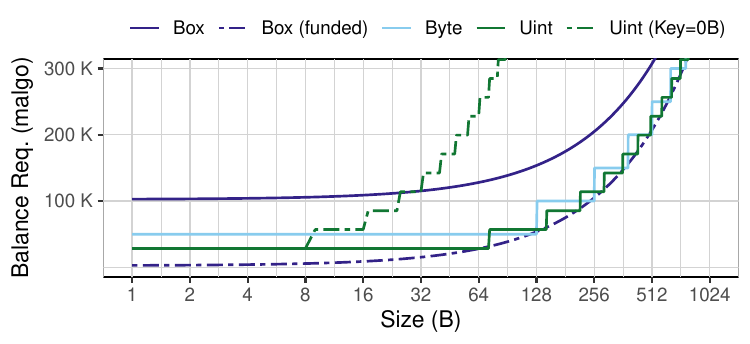}
    \vspace{-1em}
    \caption{Balance req. of Algorand storage systems as a function of their size in Byte.}
    \label{fig:mb}
    \vspace{-1em}
\end{figure}

An application architect must choose from the available storage systems. We want to investigate them from a cost perspective. Funding an account $m^{\text{acc}}_{\text{alg}}$ and deploying an application $m^{\text{app}}_{\text{alg}}$ requires a minimum balance of\footnote{Some additional cost apply to oversized contracts, see~\url{https://developer.algorand.org/docs/get-details/dapps/smart-contracts/apps\#minimum-balance-requirement-for-a-smart-contract}, last accessed 2024-06-04} 
\begin{equation}
m^{\text{acc}}_{\text{alg}} = m^{\text{app}}_{\text{alg}} = 100,000.
\end{equation}
%Beyond an application size of $2Kb$, an application requires and additional TODO for every $2Kb$ up to a limit of $8KB$.
To assess the different storage systems, we investigate their cost as a function of bytes we want to store.
We illustrate this in Figure~\ref{fig:mb}, where we show the cost in correspondence to the stored bytes up to $1024$B (with a logarithmic x-axis). The minimum balance required by the box storage must be funded by the application, thus, it requires the funding of an additional account. Consequently, box storage is only economical for larger B. However, if an application account is required and already funded (e.g., it needs to be able to send or receive payments anyway), the additional overhead does not exist.
Furthermore, if keys are merely used as indices to access storage, i.e., they don't carry any information themselves, packaging different Uint values together will reduce the balance requirements, as the Uint value itself only carries $8$B. 
%They can then be accessed by specifying an offset. The opcode cost for this functionality are negligible. 
We illustrate this by an example, and show Uint with only its value storage capacity, so with a key length of $0$B. We see that, as soon as an application requires two Uint value slots, it is more economical to package the Uint values in a Byte.
%Analogously for bytes, as soon as we require four byte values, it is more economical to package the byte values into box storage.
We use this insight in Section \ref{sec:chorpiler}.
Formally, the required minimum balance $m^{\text{s}}_{\text{alg}}$ per number of bytes $n_\text{B}$ for the different storage systems is
\begin{equation}
m^{\text{s}}_{\text{alg}}%(s)
=
\begin{cases}
28,500 \cdot (1 + \lfloor\frac{n_\text{B} - 1}{72}\rfloor) & \text{for Uint,} \\
50,000 \cdot (1 + \lfloor\frac{n_\text{B} - 1}{128}\rfloor) & \text{for Byte,} \\
m^{\text{app}}_{\text{alg}} + 2,500 + 400 \cdot n_\text{B} & \text{for Box.}
\end{cases}
\end{equation}

\noindent In summary, we can express the total balance requirement $m_{\text{alg}}$ for $n_{\text{acc}}$ amount of accounts and $n_{\text{app}}$ amount of apps as
\begin{equation}
\label{eq:mb_total}
m_{\text{alg}} = n_{\text{acc}}\cdot m^{\text{acc}}_{\text{alg}} + n_{\text{app}}\cdot m^{\text{app}}_{\text{alg}} + m^{\text{s}}_{\text{alg}}.
\end{equation}
\section{From BPMN Choreographies to Teal Applications}
\label{sec:chorpiler}
\vspaceSecAfter
%\subsection{Multi-Target Chorpiler}
In this section, we describe multi-target Chorpiler, an extension of Chorpiler's architecture introduced in Stiehle and Weber~\cite{stiehle2023processchannels}. Our aim is to facilitate quick implementation of novel blockchain targets, such as Algorand.

Since García-Bañuelos et al.~\cite{garcia-banuelosOptimizedExecutionBusiness2017a}, the state of the art in optimised code generation for smart contracts is to parse a process model into a petri net, which can then be reduced according to equivalence rules. 
In the target output, the marking of this net is then encoded as a bit array (the net is 1-safe), which allows efficient storage as an unsigned integer $i \in \mathbb{N}_0$. Then, the binary digits of $i$ correspond to all markings with tokens, e.g., $2^1$ would translate to a token in marking $M_1$.\footnote{\textit{Remark}. The number of supported places is limited by the virtual machine's supported unsinged integer size. In Ethereum, this is 256 bit; in Algorand this is 64 bit. Trivially, another integer can be allocated in Ethereum~\cite{garcia-banuelosOptimizedExecutionBusiness2017a}. As shown in Section \ref{sec:cost-model}, this is not efficient in Algorand. In Section~\ref{sec:multi_i}, we outline a technique for multi instance support using variable packaging. This technique can also be applied to increase the limitation on places.} To perform transitions, bit shifts can be executed on $i$; these can be usually performed very efficiently, and carry a negligible opcode cost. Chorpiler uses this technique to translate BPMN Choreographies to Solidity smart contracts, using a template engine. We modified Chorpiler's architecture to support multi platform targets, a goal of our architecture is a minimal core component, that assists in implementation of new blockchain targets. A specific requirement is also to support rigorous software testing, as the goal is to produce well-tested contract code. This is paramount for blockchain-based applications~\cite{lu_why_2018}. This requirement is also informed by the need of practitioners of traditional workflow execution systems, which note that this aspect is often neglected in academic artefacts \cite{ruecker2021practical}. % TODO: Section
Our separation of concerns avoids bloating the core package with blockchain platform specific dependencies, environments, and test suites. 

\begin{figure}[tb]
    \centering
    \includegraphics[width=.6\linewidth]{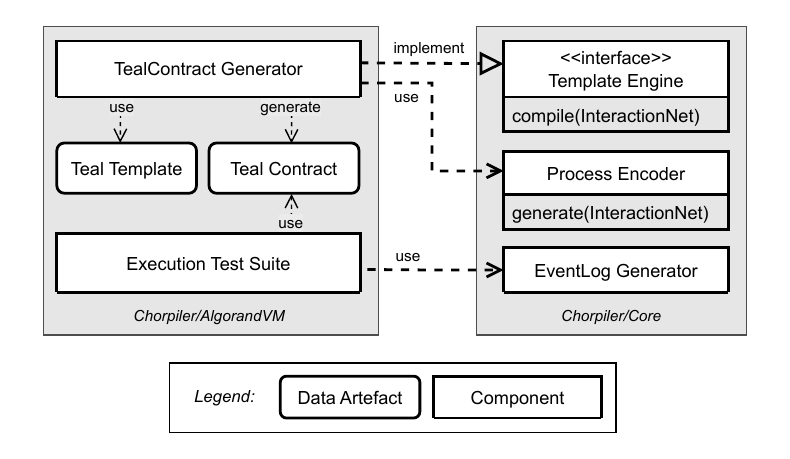}
    \vspace{-1em}
    \caption{Architecture excerpt of the refined modular Chorpiler.}
    \label{fig:architecture}
    \vspace{-1.5em}
\end{figure}

We sketch the key components of our architecture in Figure~\ref{fig:architecture}. We show the Chorpiler core package on the right side. Here, a key enabler is the introduction of a \verb|Process Encoder| component, which generates the previously described marking encoding from a petri net. A new target implementation can implement the \verb|Template Engine| interface, and can make use of the provided encoder. For the purpose of this study, we made use of this to implement a target for the Algorand virtual machine (AVM). The AVM's native language is TEAL, an assembly-like stack-oriented language. Other language bindings that compile to TEAL, e.g., PyTeal, or TealScript exist. However, PyTeal does not allow native Python expressions and TealScript is still in development and has not yet been scrutinised in security audits.
%
%We chose to forego this level of indirection and target TEAL, as it also allows more direct control. 
Using the Process Encoder architecture allowed us to implement the new target with almost no overhead. The codebase is dedicated to the back-end: the TEAL contract template and testing the produced outputs. 
Importantly, a new target implementation is not forced to use the Process Encoder, and could even use a different template engine, allowing a developer a high level of flexibility to adopt to new targets.
To facilitate testing new targets, we also introduce an \verb|EventLog Generator|. In Weber et al.~\cite{weberUntrustedBusinessProcess2016a}, the authors introduced a methodology to fuzzy test smart contract outputs for a given model. To do so, non-conforming events are generated, which are then replayed on the smart contract, which consequently must reject them.
The EventLog Generator allows the generation of such replay logs from an \textit{eXtensible Event Stream} (XES) file, containing conforming events. We make use of this in our correctness tests, which we describe in Section~\ref{sec:eval} in more detail.
%
% Teal Impl.
\vspaceSSecBefore
\subsection{Multi-Instance Execution in Teal}
\label{sec:multi_i}
\vspaceSSecAfter
To support the cost efficient interleaved execution of multiple process instances, we introduce an approach based on variable packaging, which allows a (theoretical) unlimited instantiations of the process model, without deploying a new application contract.  
We have shown in Section~\ref{sec:cost-model}, that allocating global storage slots is inefficient when the key is predictable from a certain threshold. For Uint storage, this is already for $n > 1$.

Mutli-instance execution requires the correlation of any incoming message to a receiving instance's data object~\cite{meyerAutomatingDataExchange2015}. 
For each instantiation, our design assigns a new key $i \in \mathbb{N}_0$, in an ascending order, which uniquely identifies an instance $C_i$. Here, $i$ is used as offset, to determine the storage slot in a packaged variable. Instances can be packaged into byte or box storage. Specifically, for a storage requirement of $p$ places for a process model, we require $k = 1 + \lfloor\frac{p-1}{8}\rfloor$ bytes. $C_i$ is then allocated to the next $k$ bytes from $i \times k$ in the storage up to a reasonable maximum of supported parallel instances $C_n$. 
%As we have discussed in Section~\ref{sec:cost-model}, as  $C_n \times p$ approaches $256$ byte, it becomes economical to package into box storage. 
While box storage is theoretically unlimited, a reasonable upper bound should always be provided; specifically, as box storage must be funded by the application. We provide a more detailed discussion in Section~\ref{sec:eval}.

For details on our template implementation in TEAL, we refer the interested reader to our open source artefact in Footnote~\ref{fn:artefact}.
%- teal code
%- transactional properties - how much can we draw from it?
% Multi Instance Execution
%
%\begin{figure}[t]
%    \centering
%    \includegraphics[width=.85\linewidth]{LaTeX2e+Proceedings+Templates+download/approach.pdf}
%    \caption{Approach}
%    \label{fig:enter-label}
%\end{figure}
%
%
\vspaceSecBefore
\section{Evaluation}
\label{sec:eval}
\vspaceSecAfter
To evaluate our implementation and cost models, we performed correctness and cost benchmarks on well-known process models often used in related work. Namely, the supply chain~\cite{weberUntrustedBusinessProcess2016a} (adapted from~\cite{fdhila2015changeandcompliance}) and incident management~\cite{OMG2010BPMNbyExample} case. It also allows us to compare our work to previous work on cost models, i.e, Rimba et al.~\cite{rimbaQuantifyingCostDistrust2020a} and previous work on Chorpiler, i.e, Stiehle and Weber~\cite{stiehle2023processchannels}.

We implemented a new chorpiler package as sketched in Section~\ref{sec:chorpiler}. The package is capable of generating TEAL contracts from BPMN Choreographies. The compiler supports our in Section~\ref{sec:multi_i} described multi-instance execution by packaging the AVM's \verb|uint64| types into byte or box storage.%, if that feature is required. 
We used an algorand node to deploy our contracts and simulate and execute transactions in a local sandbox test network. Our test environment runs \verb|go-algorand| in version \verb|3.24.0|, which is the latest stable release at the time of writing.\footnote{Our evaluation environment is available at the provided link in Footnote~\ref{fn:artefact}.}

We extended the chorpiler core functionalities to be able to parse XES logs and generate non-conforming traces. To benchmark the correctness of our implementation, we followed the methodology outlined in~\cite{weberUntrustedBusinessProcess2016a}.
For each case, we replayed all possible conforming traces (two for supply chain and four for the incident management case) and generated $2000$ non-conforming traces. 
To generate a non-conforming trace, Chorpiler randomly manipulates a conforming trace by a range of  operations: adding events (by evaluating other conforming events), removing an event, and swapping the position of two events. It then removes any coincidentally created conforming trace. %As a result, we replayed $1812$ non-conforming traces to the incident mgmt.\ and $1933$ to the supply chain case.}
We replayed these traces locally in our environment using the Algorand JavaScript SDK.
We performed these conformance tests on all our implemented contracts. All conforming traces were accepted and lead to the end event, and all non-conforming traces were rejected.

\begin{figure}[tb]
    \centering

    \includegraphics[width=.7\linewidth]{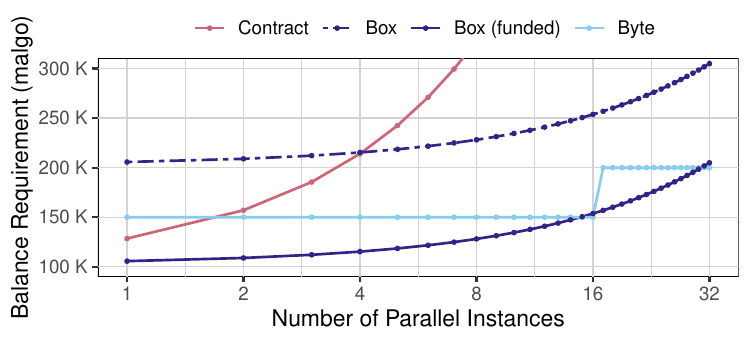}
    \vspace{-1em}
    \caption{Minimum balance requirement for multiple  parallel instance executions.}
    \label{fig:cn}
\vspace{-.5em}
\end{figure}

To assess cost, we extracted from each transaction: the opcode cost, the change in minimum balance requirement for caller and contract, and the transaction fee. We report on the results for all replayed conforming traces on both cases in Table~\ref{tab:eval}.
We show cost in the form of average opcode cost per transaction, average transaction fee per process case, and minimum balance cost per case, where the minimum balance requirement is shown exclusive of the balance requirement of the involved external accounts. We show the executions for all analysed storage systems. For both cases, a storage size of 8B, the smallest data type natively supported by the AVM, is sufficient. When considering singular case executions, Uint is unsurprisingly the most cost efficient implementation. Byte and Box contracts implement the byte packaging we introduced; however, they do not pay off and lead to higher opcode cost and balance requirements. The Box implementation also requires an additional transaction to fund the account. However, under more realistic conditions, when multiple cases are executed in parallel, the Uint implementation requires a re-deployment of the contract for every new instance, which quickly increases the minimum balance requirement. We illustrate this in Figure~\ref{fig:cn}, where we show the minimum balance requirement when multiple instances $C_n$ of supply chain or incident management are executed.
Our benchmark results closely follows our analysis in Section~\ref{sec:cost-model}, where we discussed that allocating additional Uint beyond one is inefficient. We also see that for applications with already funded application accounts (e.g., as they implement an escrow functionality), box storage is also economical for smaller $C_n$. 
We did not encounter any cost which we could not account using Equation~\ref{eq:opc} and Equation~\ref{eq:mb_total}. Notably, while we report our findings excluding the required balance to fund participant's external accounts---as they are stable costs that all implementations share---these are considerable. For the small process cases supply chain (9 participants) and incident management (8 participants), a total balance of $900\text{K}$ and $800\text{K}$ of malg is required, respectively. When including them in the cost calculation, they would amount to $85\%$ of the total execution cost of Uint.
\begin{table}[tb]
\centering
\scriptsize
\begin{tabular}{R{1.8cm}R{.2cm}R{1.1cm}R{1.1cm}R{1.2cm}R{.2cm}R{1.1cm}R{1.1cm}R{1.2cm}}
\toprule
 & & \multicolumn{3}{c}{SC ($p = 11$)} & & \multicolumn{3}{c}{IM ($p = 8$)} \\
    \cmidrule{3-5}
    \cmidrule{7-9}
 Cost & & Uint & Byte & Box & & Uint & Byte & Box \\
\midrule
(avg/tx)   op    & & $67$ & $83$ & $83$  &               & $95$ & $111$ & $110$  \\
(avg/case) tx  & & $11,000$ & $11,000$ & $12,000$ &    & $7,000$ & $7,000$ & $8,000$  \\
(case) mb  & & $128,500$ & $150,000$ & $206,100$ &   & $128,500$ & $150,000$ & $206,100$  \\

\bottomrule
\end{tabular}
\caption{Benchmark result for the generated process cases supply chain (SC) and incident management (IM), excluding external account balance requirements.}
\label{tab:eval}
\vspace{-1em}
\end{table}
\begin{table}[tb]
\centering
\scriptsize
\begin{tabular}{R{1.9cm}R{.2cm}R{1cm}R{.9cm}R{.9cm}R{.9cm}R{.9cm}}
\toprule
 &  & & \multicolumn{4}{r}{Exchange Rate in \$US} \\
 \cmidrule{4-7}
 &  & Cost & $1$ & \makecell[r]{avg.\\ 2019} & \makecell[r]{avg.\\ 2023} & \makecell[r]{avg.\\ 2024} \\ 
\midrule
 EVM & & (gas)
 \\ Chorpiler \cite{stiehle2023processchannels} & & $599,463$ & $0.020$ & $3.599$ & $35.075$ & $63.473$  \\
 %Chorpiler* & & a & a & a & a & a & a & a \\
 \midrule
 AVM & & (malg) 
 \\ (tx) & & $11,000$ & 0.011 & 0.006 & 0.002 & 0.002  \\
 (incl. mb) & & $928,500$ & 0.939 & 0.494 & 0.147 & 0.189  \\
 \midrule
SWF~\cite{rimbaQuantifyingCostDistrust2020a} & & (\$US) \\

 (24 hours)  & & $0.0009$ \\ 
 (99 years) & & $0.1816$ \\

\bottomrule
\end{tabular}
\caption{Execution cost comparison of the incident management case. We show cost in the native units: gas and malgo. We show how this translates to \$US with different exchange rates. The avg.\ prices for ETH are \$US $178.94$, $1744$, and $3156$, and for ALG $0.53$, $0.16$, and $0.20$. SWF is reported from~\cite{rimbaQuantifyingCostDistrust2020a}, with a data retention setting of one day and 99 years.}
\label{tab:compare}
\vspace{-3em}
\end{table}

We now want to put these results into context and make use of the fact that the incident management process model has been investigated in the context of process execution on Ethereum and traditional cloud computing. Comparing costs across different blockchain platforms is challenging, as it has long been understood that market capitalisation and exchange rates (e.g., ETH-US) are affected by various different market dynamics~\cite{donmez2022transaction,kukacka2023fundamental}. We follow the approach by Rimba et al., which conducted a benchmark of the Amazon Simple Workflow Service (SWF) to contrast the cost of blockchain execution~\cite{rimbaQuantifyingCostDistrust2020a}. SWF is a process-oriented software suite capable of executing workflows and offered as part of Amazon Web Services (AWS).
They report on an implementation of the incident management case and the respective cost factors. Put simply, SWF is priced based on the workflow elements executed, the execution time, and the data transfer and retention rates. We refer the reader to Rimba et al.~\cite{rimbaQuantifyingCostDistrust2020a} for more details. As they provide detailed pricing information, the recency of their results can be assessed by examining the most recent documentation provided by SWF. Indeed, we did not encounter any changes to their pricing model or technical changes rendering Rimba et al.'s results incompatible. Therefore, we deemed it unnecessary to reproduce the benchmark outlined in Rimba et al. We think it serves as a valuable illustrative baseline. And, as we show, cost of blockchain execution is still multitudes higher, likely rendering small unobserved technical changes to SWF negligible.  
We summarise our comparison in Table~\ref{tab:compare}. We show the results from Stiehle and Weber~\cite{stiehle2023processchannels}, deploying and executing the incident management process in an EVM environment. We were able to reproduce these results. For the AVM, we show the cost of the Uint implementation, as most efficient implementation for singular executions of incident management.
We show how these costs translate to \$US for various exchange rates of \$US to ETH and ALG, respectively, for the average exchange rate in 2019 (the earlierst available data point for Algorand), 2023, and 2024.\footnote{Based on the average gas price in 2023 of $33.55$ gwei, determined based on data from Etherscan (\url{https://etherscan.io/chart/gasprice}, accessed 2024-06-03). The ETH and ALG prices are the averaged daily closing prices for a given year, as reported by Yahoo Finance (\url{https://finance.yahoo.com/quote/ETH-USD/}, accessed 2024-06-03).} For SWF, we show different data retention settings. SWF fees increase when process data needs to be retained longer. In theory, blockchain data is stored indefinitely. 
Notably, for Algorand, we show transaction fees separate from balance requirements. Balance requirements include the funding of all participant's accounts. Unsurprisingly, the main driver for cost in Algorand is also storage. A driver for smaller applications, which have minimised their storage requirements through engineering efforts, is the minimum balance requirement for accounts. Specifically, for our investigated process cases, with a high ratio of participants, external account's balance requirements can make up up to 85\% (for Uint) of the total cost.
\vspaceSecBefore
\section{Discussion \& Future Directions}
\vspaceSecAfter
Comparing the execution costs of Algorand to Ethereum and to SWF, specifically for higher levels of data retention, shows the promise of process enactment on next-gen blockchains. However, the evolution of exchange rates for public blockchains is by no means stable or predictable. We illustrated the steep climb of ETH from 2019 on---calculations which are in line with previous work~\cite{stiehle2022blockchain}. Algorand has charted a somewhat reverse trend. Both platforms exhibit opposing qualities in many other aspects. While Ethereum utilises a gas fee market, Algorand defines fixed fee costs. While this increases predictability and usability in the immediate sense, should Algorand rise in popularity as a speculative asset, the fixed fee nature would force protocol changes to adapt to a changed exchange rate. Currently, Algorand facilitates important protocol changes by a governance model, where votes are open for a period of three months.\footnote{\url{https://governance.algorand.foundation}, accessed 2024-06-04.} In many ways, Algorand has not yet come under such scrutiny as Ethereum, whose operators were forced to make many adjustments to its cost model based on discovered or exploited attack vectors in the past (see e.g.,~\cite{perez2019broken}). With EIP-1559, Ethereum reduced the level of user agency in the fee market in favour of better usability. Furthermore, certain market dynamics have introduced unintended consequences, which forced drastic reductions in gas refunds~\cite{buterin2021eip3529}. 
Conversely, the minimum balance system of Algorand operates independently of the transaction fee system. While we treated minimum balance requirements as cost, they essentially resemble invested or frozen assets, where notions like opportunity cost apply. The balance requirement is also exposed to devaluation and appreciation effects as the exchange rate of ALG to other currencies changes. As we have shown, minimum balance requirements can constitute for up to 85\% of the total cost, so these considerations are paramount when considering Algorand.
Finally, while Ethereum incentivizes honest verifiers by rewards, Algorand does not currently distribute rewards based on blocks or transactions. Rather, rewards are distributed by so-called reward programs.\footnote{\url{https://www.algorand.foundation/200-million-algo-staking-rewards-program}, accessed 2024-06-04.} 

Our results show the promise of blockchain-based process execution on next-gen blockchains. However, our study also stands as early evidence that adapting existing approaches to novel protocols is by no means trivial. Protocol designs can differ greatly, and involve novel trade-offs, implications, and attack vectors. Superficial understanding of new protocols can in principle lead to poor design decisions and in consequence to security vulnerabilities, wasted effort, and cost.
%Many unknowns remain, from both an academic and practitioner perspective. 

Highlighting the above design differences illustrates the need for a systematic investigation of the design space. Conducting a kind of design archaeology~\cite{chandra2019design}, and unearthing different design principles as well as constructing taxonomies, could bring immense value to both users and designers of decentralised protocols---not just in the design of such systems, but also in communicating design decisions~\cite{williamsDesignEmergingDigital2008}.

While we have illustrated how concrete cost models can influence the design of an approach \textit{a priori}, and help weigh different design options, the AVM results shown are all based on static local experiment runs. We see opportunity in automated simulation approaches to assess existing implementations in more realistic environments that are capable to simulate e.g., different levels of congestion. Simulation can facilitate fast and secure redesign of different process implementations~\cite{pufahl2018design}. In the case of large decentralised systems, this constitutes a major challenge. Specifically, studying systems under load in an economical manner remains a challenge. Current benchmark systems~\cite{rezabek2023multilayer,gramoli2023diablo} focus on the system as a whole. They do not natively support reactive applications like choreographies, where participants observe the blockchain, and execute contracts as a consequence of observed events.
%Tested various edge cases.
%Average gas cost in 2023 = 33.54445 gwei (etherscan) = 3.354445e-8 Ether mal 599463 = 0.02010865663. 
%As comparison, we used Chorpiler's existing translation capabilities to (TODO: mention that the core still is meant to do EVM) generate the same models. 
%Compared to Stiehle and Weber~\cite{stiehle2023processchannels}, we introduced some optimisation techniques, such as enabling the use of the most recent Solidity compiler optimisations, and prioritising cheaper operations, to reduce the search space first, and delaying more costly operations.
%We use models commenly known from previous work, to facilitate comparison. Specifically, AWS
%
\vspaceSecBefore
\section{Conclusion}
\vspaceSecAfter
Blockchain-based process systems must deal with platform specific limitations, cost models, and security considerations. This is especially pronounced in blockchain, where design knowledge is not mature. 
We have introduced a compiler architecture, which supports implementing new blockchain targets with minimal overhead, and demonstrated its use for the Algorand blockchain. We presented a comprehensive analysis and formalisation of the Algorand cost model. Our presented approach, a method to compile BPMN choreographies to TEAL contracts, makes immediate use of these insights.
Our manuscript illustrates that assessing cost for novel blockchain protocols is by no means trivial. However, for organisations, predictable cost and performance estimates are paramount---without an understanding of operational cost, conducting a cost-benefit analysis is difficult if not impossible.
We see our study as a contribution to a more holistic understanding of novel blockchains.

%An often overlooked aspect, testability, has been incorporated to facilitate correctness benchmarks. Model-driven solutions for blockchain-based process systems have been investigated at large by the community. Other initiatives beyond process underline this interest\footnote{https://wizard.openzeppelin.com}.
%However, these capabilities must be strengthened and extended in future work. For example, novel fuzzing and verification approaches could be adopted 

%While we have illustrated how concrete cost models can influence the design of an approach and help weigh different design options, the results are limited to static local experiment runs. We see opportunity in automated simulation approaches. Simulation can facilitate fast and secure redesign of different process implementations~\cite{pufahl2018design}. In the case of large decentralised systems, this constitutes a major challenge. Specifically, studying systems under load in an economical manner remains a challenge. Current benchmark systems~\cite{rezabek2023multilayer,gramoli2023diablo} focus on the system as a whole. They do not natively support reactive applications like choreographies, where participants observe the blockchain, and execute contracts as a reaction to the observed.
\vspace{-.6em}
\credits{
\subsubsection*{\ackname}
The authors wish to disclose the use of generative AI to assist in the editing of the manuscript and the implementation of the artefacts.
Generated output was never taken "as-is", it was reviewed and verified by the authors.}
\bibliographystyle{splncs04}
\bibliography{bib}
\end{document}